\documentclass[useAMS,usenatbib]{mn2e}
\usepackage{txfonts}
\usepackage{graphicx}
\usepackage[mathscr]{eucal}
\newcommand{\be}{\begin{equation}}
\newcommand{\ee}{\end{equation}}
\newcommand{\bdm}{\begin{displaymath}}
\newcommand{\edm}{\end{displaymath}}
\def\pdot {\dot P}
\def\nudot {\dot \nu}

\def\msun{{\rm M}_{\odot}}
\def\rsun{R_{\odot}}

\def\mdot {\dot M_{\rm W}}

\def\ltsima{$\; \buildrel < \over \sim \;$}
\def\lsim{\lower.5ex\hbox{\ltsima}}
\def\gtsima{$\; \buildrel > \over \sim \;$}
\def\gsim{\lower.5ex\hbox{\gtsima}}
\def\hd {HD\,49798}
\def\rx {RX J0648.0--4418}
\def\hr {HD\,49798/RX\,J0648.0--4418}
\def\bd     {BD\,+37$^{\circ}$442}
\def\xmm {\emph{XMM-Newton}}

\def\swi {\emph{Swift}}

\title[Spin-up of  HD 49798 companion]{Discovery of spin-up in the X-ray pulsar  companion of the hot subdwarf HD 49798}
\author[Mereghetti S. et al.]{Sandro Mereghetti$^{1}$\thanks{E-mail:
sandro@iasf-milano.inaf.it}, Fabio Pintore$^1$, Paolo Esposito$^{1}$, Nicola La Palombara$^{1}$, 
  \newauthor  Andrea Tiengo$^{1,2,3}$,  Gian Luca Israel$^4$,  Luigi Stella$^4$  \\
    $^1$ INAF -- IASF Milano, Via E. Bassini 15, I-20133 Milano, Italy\\
   $^2$ Istituto Universitario di Studi Superiori, Piazza della Vittoria 15, I-27100 Pavia, Italy\\
  $^3$ Istituto Nazionale di Fisica Nucleare, Sezione di Pavia, Via A. Bassi 6, I-27100 Pavia, Italy\\  
    $^4$ INAF -- Osservatorio Astronomico di Roma, Via Frascati 33, I-00040, Monte Porzio Catone, Italy}

\begin{document}
\date{Accepted 2016 March 2; Received 2016 March 2; in original form 2015 December 30}
\pagerange{\pageref{firstpage}--\pageref{lastpage}}
\pubyear{}
\maketitle
\label{firstpage}
\begin{abstract}
The hot subdwarf   \hd\ has an X-ray emitting compact companion with a spin-period of 13.2 s and a dynamically measured mass of 1.28$\pm$0.05 $\msun$, consistent with either a neutron star or a white dwarf. Using all the available \xmm\ and \swi\ observations of this source, we could perform a phase-connected timing analysis extending back to the $ROSAT$ data obtained in 1992.  We found that the pulsar is spinning up at a rate of (2.15$\pm$0.05)$\times10^{-15}$ s  s$^{-1}$.  This result is best interpreted in terms of a neutron star accreting from the  wind of its subdwarf companion, although the remarkably steady period derivative over more than 20 years is unusual in wind-accreting neutron stars. The possibility that the compact object is a massive white dwarf accreting through a disk cannot be excluded, but it requires a larger distance   and/or   properties of the stellar wind of \hd\ different  from those derived from the modelling of its optical/UV spectra.
\end{abstract}


\section{Introduction}

\hr\ is a binary system with   orbital period of 1.55 days containing an  X-ray pulsar with   spin period of 13.2 s \citep{tha70,sti94,isr97,mer11}.  
This X-ray source  is particularly interesting since it is the only known   pulsar with a hot subdwarf   companion\footnote{a possibly similar system, \bd , has been reported by  \citet{lap12}, but the detection of an X-ray periodicity awaits confirmation.}.
Its properties are quite different from those of all the other   X-ray binaries containing accretion-powered neutron stars or white dwarfs.
The  X-ray emission from \rx\ shows two distinct features: a  strongly pulsed soft thermal component, well fit by  a blackbody of temperature kT$\sim$30 eV,  and a harder tail (power law photon index $\sim$2), which accounts for most of the emission above 0.6 keV.
An upper limit  on the pulsar spin period derivative of $|\pdot|<6\times10^{-15}$ s s$^{-1}$ (90\% c.l.) was derived by phase-connecting \xmm\ observations obtained in 2008 and 2011 \citep{mer13}.

The masses of the two stars in this binary system are well constrained by the measurement of the optical and X-ray mass functions, the system inclination being derived from the duration of the X-ray eclipse \citep{mer09}:
with a mass of   1.50$\pm$0.05 $\msun$, \hd\ is among the most massive hot subdwarfs of O spectral type, while the mass of the pulsating X-ray source  is 1.28$\pm$0.05  $\msun$.   

\hd\ is one of the few hot subdwarfs for which evidence for a stellar wind has been reported \citep{ham81,ham10}.
It was thus natural to explain the relatively small X-ray luminosity of this binary (see Sect.~\ref{sec_lum})  as emission powered  by wind accretion onto  either a white dwarf (WD) or a neutron star (NS).
The WD hypothesis was favored by  \citet{mer09}, based on a simple estimate of the expected luminosity resulting from the wind parameters reported by \citet{ham81} and assuming a distance of 650 pc \citep{kud78}.

To better understand this system, we have performed a timing analysis of all the available X-ray observations in which the 13.2 s pulsations can be detected with a high confidence level. This has allowed us to derive a phase-connected timing solution and to measure for the first time the pulsar spin-up (Section \ref{analysis}). On the basis of this result, in Section \ref{discussion}  we  rediscuss  the nature of the compact companion of \hd .

\begin{table} 
\caption{Log of the \xmm\ observations of \rx.
\label{table_obs}}
\begin{tabular}{ccccc}
\hline
   Observation & Start time &  End  time & Duration  \\
   date & MJD  & MJD  &  pn/MOS (ks) \\\hline
2002 May 03   &   52397.46    & 52397.55 &    4.5 / 7.2    \\ 
2002 May 04   &   52397.98    & 52398.06 &     1.4 / 5.6    \\
 2002 May 04   &   52398.56    & 52398.59 &    0.6 / 2.5     \\
 2002 Sep 17   &   52534.58    & 52534.72 &    6.9 / 11.9     \\
2008 May 10   &   54596.90    & 54597.38 &   36.7 / 43.0     \\
2011 May 02   &   55683.55  &   55683.76  &   17.0 / 18.5     \\
2011 Aug 18  &   55791.88  &   55792.07  &   15.0 / 16.6       \\
2011 Aug 20  &   55793.46  &   55793.62  &  11.8 / 14.3         \\
2011 Aug 25  &   55798.04  &   55798.27  &   18.0 / 19.6         \\
2011 Sep 03   &   55807.35  &   55807.54  &  15.0 / 16.6           \\
 2011 Sep 08  &   55811.10  &   55812.19  &   15.0 / 16.6           \\
2013 Nov 10  &   56605.80 &     56606.25 &   37.9 / 39.5            \\
 2014 Oct 18  &   56948.37   &   56948.71 &  27.1 / 29.4              \\
\hline
\end{tabular}
\end{table}

 \begin{figure*}
 \begin{center}
\includegraphics[width=12.cm,angle=0]{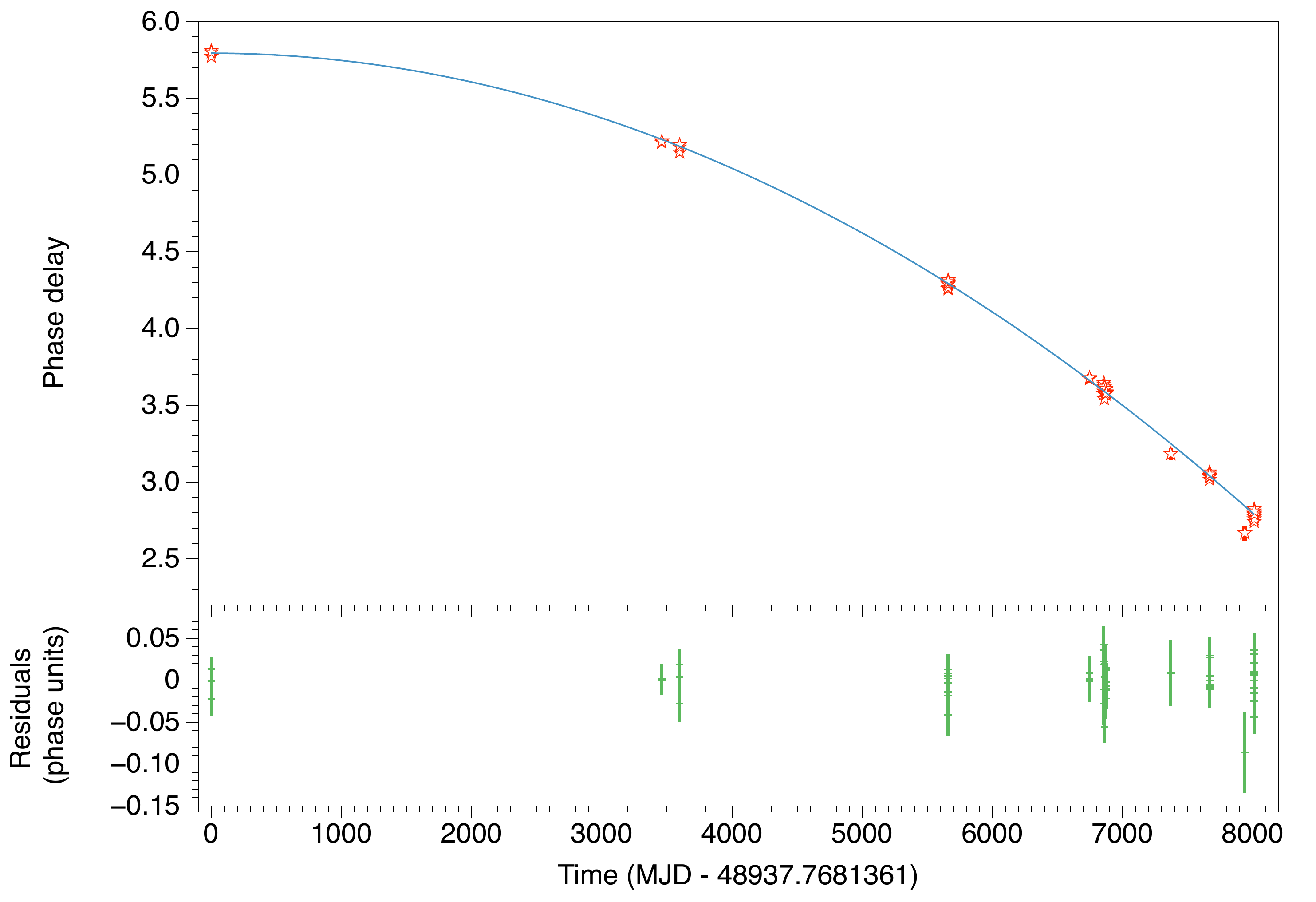}
\caption{Phases of the pulses (in units of cycles) fitted with a quadratic function (top panel) and residuals from the best fit (bottom panel) .      }
 \label{fig_phasefit}
 \end{center}
 \end{figure*}

\section{Observations and data analysis}
 \label{analysis}
 
Since  May 2002 to  October 2014, several observations of \hd\  were carried out using the  \xmm\ satellite  (see Table \ref{table_obs}). 
In this work we used the data collected with the three CCD cameras of the EPIC instrument. 
In all the observations  they were operated in Full Frame mode, which provides a time resolution of 73 ms for the pn camera and of 2.6 s for the two MOS cameras.
Source  counts were  extracted from a circle of 30$''$ radius around the  position of \hd . 

\hd\ was also observed with the ROSAT satellite on November 11,  1992. These are the data in which the pulsations were discovered \citep{isr97}, and they allow us to extend the baseline of our phase-connected timing solution.   For  the analysis reported here we used the counts obtained with the PSPC instrument in the 0.1--0.5 keV energy range (extraction radius of 1 arcmin). 
Finally, we included in the timing analysis for the phase-connection also  the observations carried out with the \swi\  $XRT$ instrument in   January 2013,  May 2014,   August 2014,  and January/February 2015.

\subsection{Timing analysis}
\label{sec_tim}

The times of arrival were converted to the Solar system barycenter and corrected for the orbital motion using the parameters given in Table  \ref{table_ephem}. The phases of the pulsations were derived by fitting a constant plus a sinusoid to the folded pulse profiles. 
For the EPIC data we  used the pn profiles in the energy range 0.15--0.5 keV,  which have  the largest pulsed fraction.
In order to obtain a phase-connected timing solution, we fitted the pulse phases as a function of time with the function

\begin{equation}
\phi(t) = \phi_0 + \nu_0 (t-T_0) + 0.5 \dot\nu (t-T_0)^2
\end{equation}
 
\noindent
where $\nu=1/P$ is the star rotation frequency.
We started by using the most closely spaced observations (Summer 2011) and gradually included in the fit the other observations, as increasingly more precise fit parameters allowed us to keep track of the number of intervening pulses.    In this way we could obtain a unique solution giving a  good fit to all the considered data with the values of $\nu$, $\nudot$ and orbital parameters reported in Table  \ref{table_ephem}. The fit to the data points and the residuals are shown in Fig.~\ref{fig_phasefit}.   
The highly statistically significant  quadratic term in the fit corresponds to a spin-up rate  of    $\pdot=(-2.15\pm0.05)\times10^{-15}$ s s$^{-1}$.

\begin{table}
\caption{Parameters of the binary system \hr .
\label{table_ephem}}
\begin{tabular}{lcl}
\hline
 Parameter  &  Value &   Units   \\
\hline
Right Ascension & 6$^h$ 48$^m$ 4.7$^s$  &  J2000  \\
Declination    &   $-44^{\circ}$ 18$'$ 58.4$''$   &  J2000 \\
Orbital period  & 1.547666(2)              &  d      \\ 
A$_X$ sin i    &   9.79(19)                &  light-s      \\
T$^*$        & 43961.243(15)            &  MJD \\
$\nu_0$       &    0.0758480846(1) & Hz \\
$\nudot$     &    1.24(2) $\times$ 10$^{-17}$               &  Hz s$^{-1}$    \\
P$_0$          &   13.18424856(2)  & s \\
$\pdot$     &   --2.15(5)  $\times$ 10$^{-15}$          &  s s$^{-1}$ \\
T$_0$    &  48937. 7681361 & MJD \\
M$_X$    &  1.28(5)  & $\msun$ \\
M$_C$    &  1.50(5)  & $\msun$  \\
\hline
\end{tabular}
\begin{small}
\\
\\
\end{small}
\end{table}

\begin{figure}
 \begin{center}
\includegraphics[width=15.cm,angle=0]{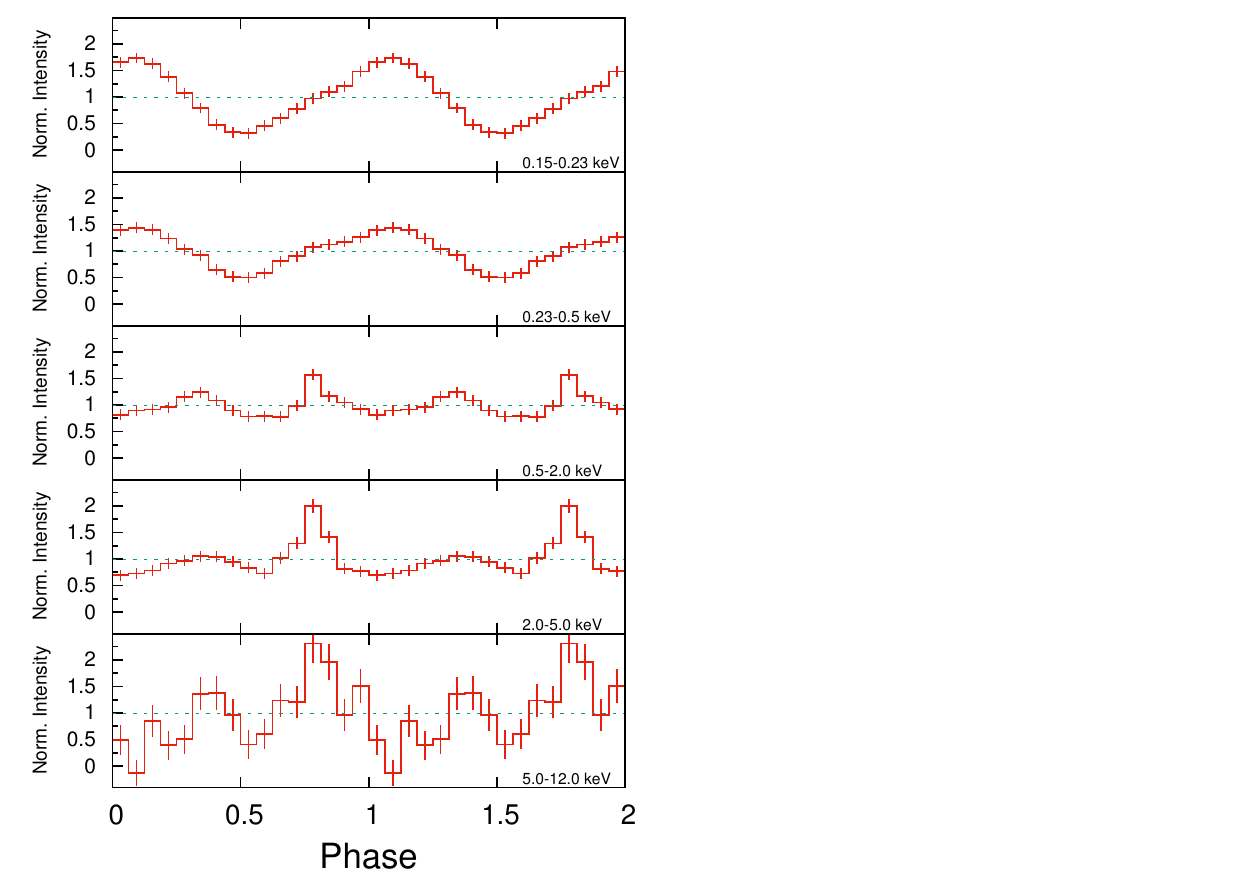}
\caption{Folded light curves in different energy ranges obtained by summing all the \xmm\ observations (pn+MOS). The estimated background contribution has been subtracted and the profiles have been normalized to the average count rate.  }
 \label{fig_lcurves}
 \end{center}
 \end{figure}

Based on the phase-connected timing solution we could join all the \xmm\ observations  and  produce the background-subtracted pulse profiles in different energy ranges  shown in Fig.~\ref{fig_lcurves}.
This figure shows a striking change in the pulse profile at an energy of 0.5 keV.
At lower energies, where the thermal spectral component dominates, the light curves are characterized by a single broad pulse. The pulsed fraction (defined as the amplitude of a fitted sinusoid divided by the average count rate) is 65$\pm$1\%  in the 0.15-0.23 keV  range and 44$\pm$1\%  in the 0.23-0.5 keV  range. 
Above 0.5 keV the light curve is more complex, displaying two pulses not aligned in phase with the low-energy one and with an energy-dependent relative intensity.

\subsection{Spectral analysis and luminosity}
\label{sec_lum}

In this subsection we reassess  the total accretion-powered  luminosity of the  \hd\ companion, exploiting all the available data and taking into  account the uncertainties in the  spectral parameters derived from the  fits. 
This is particularly important since, due to the very soft spectrum, a large fraction of the luminosity might be emitted at energies smaller than $\sim$0.2 keV,  below the observed X-ray range. 
X-rays from this system are also detected during the eclipse. They are most likely emitted in the wind of \hd\ and it is reasonable to assume that they are present during the whole orbital revolution.  
Therefore, it is necessary to model this component  and subtract it from the non-eclipsed spectrum in order to  properly estimate the bolometric luminosity of the X-ray pulsar.
 
Hence, we first analyzed all the   \xmm\ data  in which the  eclipse was present (8 observations in  2008, 2011 and 2013). 
We stacked together all the EPIC-pn data of the eclipses for a total exposure of $\sim$30 ks, and we did the same for the EPIC-MOS1 and MOS2 data.
We fitted together the resulting pn and MOS spectra with the sum of three {\sc Mekal} components \citep{mew85} with  temperatures of 0.1, 0.6 and 4 keV and abundances fixed at those of \hd\  \citep{mer15}. 
This model was then included as a fixed component in all  the fits of the non-eclipsed X-ray emission described below.

\begin{figure}
 \begin{center}
\includegraphics[width=6.5cm,angle=90]{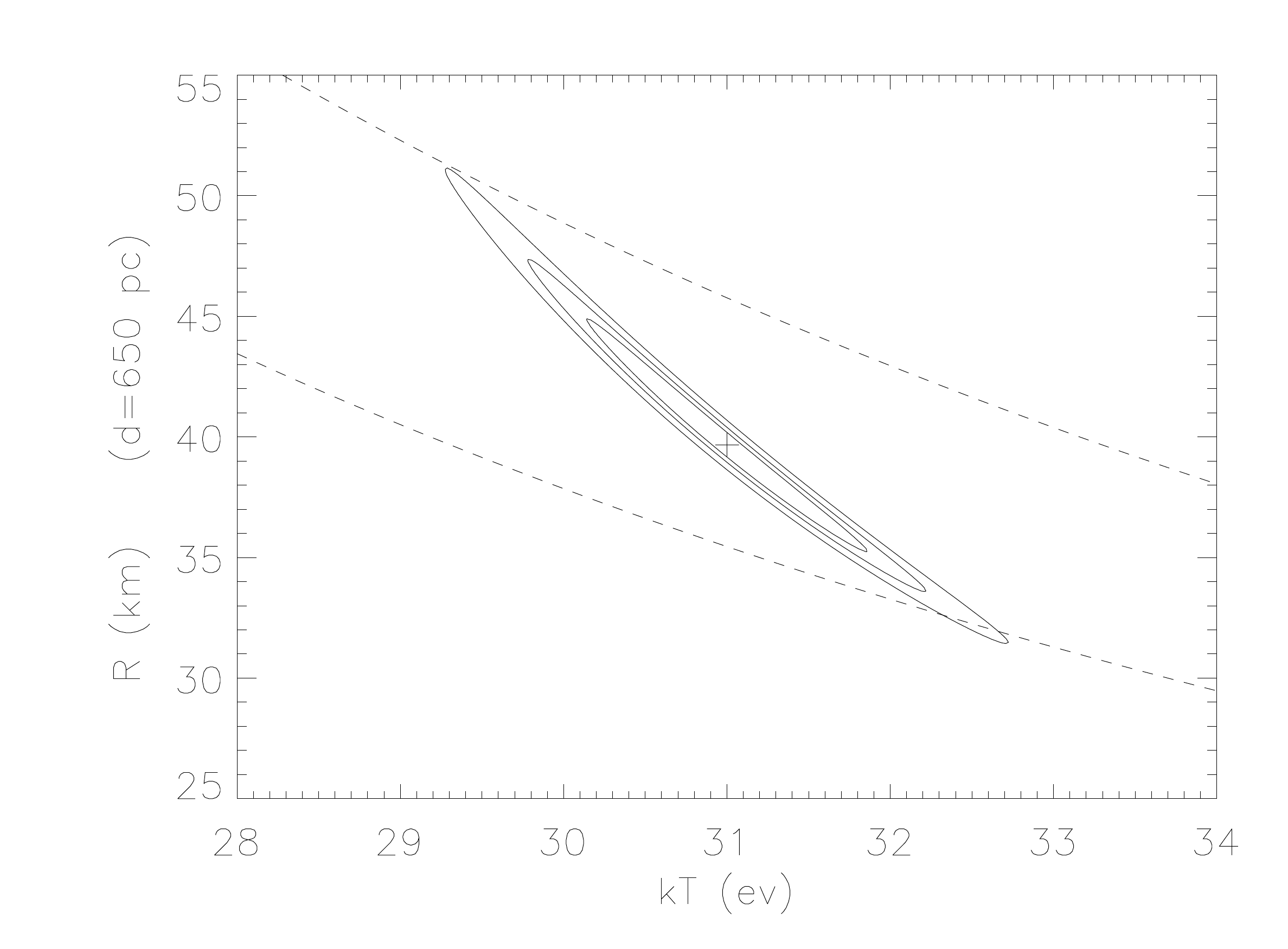}
\caption{Confidence level contours  (1, 2 and 3 $\sigma$) for the blackbody temperature and radius of emitting area (assuming a distance of 650 pc). The two dashed lines correspond to bolometric luminosities of $2.5\times10^{32}$  erg s$^{-1}$ (upper curve) and $1.5\times10^{32}$  erg s$^{-1}$ (lower curve). }
 \label{fig_cont}
 \end{center}
 \end{figure}

We then summed all the  \xmm\ data (from 2002 to 2014) excluding the time intervals corresponding to the pulsar eclipse. This resulted in a total exposure of $\sim90$ ks.  These stacked spectra (one for each camera) were fitted  with  a blackbody  plus powerlaw model,  including as a fixed component the eclipse model.
The resulting  best-fit  values of the spectral parameters are given in Table~\ref{tab_spectra}.
Varying the model parameters within their  $3\sigma$ c.l. regions  (see Fig.\ref{fig_cont}), we found that the bolometric flux of the blackbody component is constrained within the range $(3-5)\times10^{-12}$ erg cm$^{-2}$ s$^{-1}$. 
The  flux of the power-law  component depends on the considered energy range, but it is at most   $\sim2\times10^{-13}$ erg cm$^{-2}$ s$^{-1}$  (corrected for the interstellar absorption), even considering the wide energy range  0.01-10 keV.  Therefore, the power-law component gives  only a minor contribution to the total flux and we can conclude that the total accretion-powered luminosity, $L_X$, is in the range $(2\pm0.5)\times10^{32}~({\rm d}/{\rm 650 ~pc})^2$ erg s$^{-1}$.  

 \section{Discussion}
 \label{discussion}

Our  discovery  of a long term  spin-up confirms that the pulsar companion of \hd\ is accreting mass. 
 From the observed value   $\nudot=1.24\times10^{-17}$ Hz s$^{-1}$,  we can   estimate the amount of specific angular momentum which is accreted by the compact object, which, using the relation $L_X = GM\dot M/R$, can be written as:

\begin{equation}
j =  \frac{2 \pi \nudot I G M} {L_X R},
\label{eq_j}
\end{equation}

\noindent
where $M$, $R$ and $I$ are the mass, radius, and moment of inertia of the compact object and $\dot M$ is the mass accretion rate. 
In the case of a  WD  (for which we assume $I=10^{50}$ g cm$^2$, $R=3000$ km, $M=1.28 ~\msun$) we obtain

\begin{equation}
j_{WD} =    2.2\times10^{19} \left({L_X}\over{2\times10^{32}~{\rm erg~s^{-1}}}\right)^{-1}  ~{\rm cm^2~s^{-1}} , 
\label{eq_j_wd}
\end{equation}

\noindent
while for a  NS  ($I=10^{45}$ g cm$^2$, $R=12$ km, $M=1.28 ~\msun$) we have:

\begin{equation}
j_{NS} =  5.5\times10^{16} \left({L_X}\over{2\times10^{32}~{\rm erg~s^{-1}}}\right)^{-1}  ~{\rm cm^2~s^{-1}}.
\label{eq_j_ns}
\end{equation}

\noindent
These values should be compared with the specific angular momentum that can be provided by the inflow of the accreting matter. Estimates of the specific  angular momentum in the case of wind accretion are subject to many uncertainties because the properties of the   flow of the matter gravitationaly captured by the compact object moving through the wind of the companion are not well known. Estimates have been made by different authors \citep[see, e.g.,][]{ill75,sha76,wan81,ruf97} under some simplyfing hypothesis which might not apply in our case. However, we can take as a reasonable upper limit the value $j_w  \sim \Omega R_G^2$,   where $\Omega$ is the orbital angular velocity and $R_G$  is the  Bondi-Hoyle accretion radius ($R_G =   2 G M / V_{REL}^2$, where  $V_{REL}$ is the relative velocity between the compact object and the stellar wind). In the case of \hd\ we have :

\begin{equation}
j_w =     5.4\times10^{16} \left({V_{REL}}\over{1000 ~{\rm km~s^{-1}}}\right)^{-4}  ~{\rm cm^2~s^{-1}}   
\label{eq_j_w}
\end{equation}

\noindent
which, compared to eq.~\ref{eq_j_wd},   shows that, in this simple wind-accretion scenario,  it is difficult to obtain the observed spin-up rate if the compact object is a WD. 

The case of a wind-accreting NS  is more favourable, but still a rather small value of  $V_{REL}$ is necessary to provide enough angular momentum (see eqs.~\ref{eq_j_w} and ~\ref{eq_j_ns}). Furthermore, a small value of $V_{REL}$ might be inconsistent with the observed luminosity. In fact 
the accretion rate can be estimated from the relation 

\begin{equation}
\dot M =  \mdot \left({R_G}\over{2a}\right)^2  
\label{eq_mdot}
\end{equation}

\noindent
where   $a$ is the orbital separation and $\mdot$   the wind mass-loss rate from the companion star. 
The properties of the wind of \hd\ have been derived by  \citet{ham10}, who obtained  $\mdot=3\times10^{-9}$ $\msun$ yr$^{-1}$ = $2\times10^{17}$ g s$^{-1}$ and an  asymptotic wind velocity $V_{\infty}=1350$ km s$^{-1}$.  
The   luminosity   corresponding to eq.~\ref{eq_mdot} is

\begin{equation}
L_{NS} =  1.3\times10^{34}~\left( {\mdot} \over {{\rm 10^{17}~g~s^{-1}}} \right) \left( {V_{REL}} \over {{\rm 1000~{\rm km~s^{-1}}}} \right)^{-4}    ~~{\rm erg~s^{-1}},     
\label{L_ns}
\end{equation}

 \noindent
which is much larger than the observed value.

In view of these difficulties, we consider in the following  the case of disk accretion.
If the accretion occurs through a Keplerian disk,  the  spin-up rate is given by:

\begin{equation}
2\pi I \nudot  = \dot M \sqrt{G M R_M}.
\label{eq_torque}
\end{equation}

\noindent 
$R_M$ is the magnetospheric radius, which can be expressed  by 

\begin{equation}
R_M = \xi  \left({ \mu^4 }\over{G M \dot M^2}\right)^{1/7},
\label{eq_alfven}
\end{equation}

\noindent
where $\mu$ is the magnetic dipole moment  ($\mu = B R^3/2$) 
and $\xi$ a parameter of the order of unity.
 
Using these equations, we can derive the minimum magnetic field $B$ required to provide the observed spin-up rate.
The minimum field  is different in the case of a WD and of a NS:

\begin{equation}
B_{WD} >  1.4\times10^{9} ~ \xi^{-7/4}  \left(   {\dot M} \over {{\rm 10^{15}~g~s^{-1}}} \right)^{-3}  ~~~~~~{\rm G}      
\label{eq_B_WD}
\end{equation}

\begin{equation}
B_{NS} >  6.9\times10^{10} ~ \xi^{-7/4}  \left(   {\dot M} \over {{\rm 10^{11}~g~s^{-1}}} \right)^{-3}  ~~~~~~{\rm G}.    
\label{eq_B_NS}
\end{equation}

In order to have accretion and spin-up,   $R_M$ must be smaller than the corotation radius, $R_{COR}= (G M P^2 /4\pi^2)^{1/3} = 9\times10^8$ cm. This condition yields  an upper limit on B, which again is different in the WD and NS case:

\begin{equation}
B_{WD} <  4\times10^{4} ~ \xi^{-7/4}  \left(   {\dot M} \over {{\rm 10^{15}~g~s^{-1}}} \right)^{1/2}  ~~~~~~{\rm G}      
\label{eq_Bmax_wd}
\end{equation}

\begin{equation}
B_{NS} <  6.3\times10^{9} ~ \xi^{-7/4}  \left(   {\dot M} \over {{\rm 10^{11}~g~s^{-1}}} \right)^{1/2}  ~~~~~~{\rm G}.    
\label{eq_Bmax_ns}
\end{equation}

\noindent
The above requirements for a minimum (eqs. \ref{eq_B_WD} and \ref{eq_B_NS}) and a maximum magnetic field (eqs. \ref{eq_Bmax_wd} and \ref{eq_Bmax_ns}) can be satisfied only if

\begin{equation}
\dot M > 2\times10^{16}  ~~~{\rm g~s^{-1}, ~~~~~in~ the~ case~ of~ a~ WD,}
\label{eq_Lmin_wd}
\end{equation}

\begin{equation}
\dot M > 2\times10^{11}  ~~~{\rm g~s^{-1}, ~~~~~in~ the~ case~ of~ a~ NS}.
\label{eq_Lmin_ns}
\end{equation}

We can thus conclude that, in case of disk accretion, it is possible to obtain the observed spin-up rate both in the WD and NS case, provided that the  accretion rate is   large enough.  In the next subsection we will discuss whether  this requirement is consistent with the observed X-ray flux and the implications for the nature of the compact object.

\begin{table}
\caption{Spectral results  (errors at 90\% c.l. for a single parameter)  
\label{tab_spectra}
}
\begin{tabular}{lcl}
\hline
Parameter                           & Value          & Units \\
\hline
$N_{\mathrm{H}}$              & $<8\times10^{20}$          &  cm$^{-2}$		           \\
$kT_{\rm BB}$                    & 31$\pm$1                 	 & eV     \\
$R_{\rm BB}^{a}$ 		   	& $39.6_{-4.8}^{+5.7}$	  & km     \\
Photon index		 	  & 1.88$\pm$0.08	   &   \\
$K_{\rm PL}^{b}$            & (1.75$\pm$0.08)$\times$10$^{-5}$ & 	ph cm$^{-2}$ s$^{-1}$ keV$^{-1}$	    \\
\hline
\end{tabular}
\begin{small}
\\
$^{a}$ Blackbody emission radius for $d=650$ pc.
\\
$^{b}$  Power law flux at 1 keV.
\\
\end{small}
\end{table}

\subsection{Nature of the compact companion star}

As discussed above, in the case of a WD companion it is very difficult to obtain the observed  spin-up rate, unless an    accretion disk is formed. 
This is more likely to happen if  the mass donor is (nearly) filling its Roche-lobe. 
The Roche-lobe in this system has a radius  $R_{RL}$$\sim$3  $\rsun$.  \citet{kud78} estimated that the radius of  \hd\ is $R_{HD}$= 1.45$\pm$0.25  $\rsun$, based on a distance $d$=650 pc. However,  a larger distance is required in the WD interpretation (see below) and, considering that the  estimated radius scales linearly with $d$, it is then possible that $R_{HD}\sim R_{RL}$. 

The constraints on the magnetic field strength (eqs.~\ref{eq_B_WD} and \ref{eq_Bmax_wd}) are compatible with the   values observed in weakly magnetized WDs \citep{fer15}.
For example, a magnetic field of  $\sim$200 kG, would be sufficiently large to produce the required torque, yet giving $R_M<R_{COR}$. 
However, the luminosity corresponding to the  minimum mass accretion rate given by eq. \ref{eq_Lmin_wd} can be reconciled with the value  derived in section \ref{sec_lum} only if  the source  distance  is larger than $\sim$4  kpc  (unless the radiative efficiency of the accretion flow is particularly small  or most of the accretion-powered luminosity is emitted in the unobserved far UV range).  

The distance of 650 pc, commonly adopted in the literature, was derived from the equivalent width of the Ca  interstellar lines  in the spectrum of \hd\ \citep{kud78} and, considering the Galactic latitude of this star ($b=-19^{\circ}$), it could be underestimated. The revised parallax of \hd ,  obtained with Hipparcos  (1.2$\pm$0.5  mas), corresponds to a distance of 830 pc.  However,  this parallax measurement has a relatively small statistical significance and a distance  up to 5 kpc is  within the  2$\sigma$ error.

If the compact companion of \hd\ is  a NS, the condition on the minimum accretion  rate derived from the torque requirement (eq.~\ref{eq_Lmin_ns}) predicts an X-ray luminosity $L_X>3\times10^{31}$ erg s$^{-1}$,     consistent with the observations. The NS dipole   field is  constrained by  eqs.~\ref{eq_B_NS} and \ref{eq_Bmax_ns}, which, for example,  lead to  $2\times10^7<B_{NS}<3\times10^{10}$ G for   $\dot M=1.5\times10^{12}$ g s$^{-1}$   ($L_X=2\times10^{32}$ erg s$^{-1}$). 
This field is rather low when compared to those usually found in accreting pulsars in high mass X-ray binaries \citep[e.g.,][]{rev15} and  might be related  to the particular evolution leading to a NS with a hot subdwarf companion \citep{ibe93}. 
Another remarkable aspect of this X-ray binary is the stability of the spin-up rate, mantained  for a time period longer than 20 years. 
All the known accreting neutron stars in (high-mass) X-ray binaries show  variations in their spin-period derivative \citep[see, e.g.,][and references therein]{chak97} and, to our knowledge, for none of them it has been possible to obtain phase-connected timing solutions spanning such a long time interval of steady spin-up.

Finally, we note that the large radius of the emitting area derived from the blackbody spectral fit,  $R_{BB} \sim 40~(d/650~{\rm pc})$ km, is also puzzling. Even allowing for the uncertainties in the best-fit parameters (Fig.~\ref{fig_cont}), this value is larger than the radius of a NS, while the 
high pulsed fraction    suggests that the  thermal emission originates  from a relatively small fraction of the star. 

\section{Conclusions}

By phase-connecting \xmm  , $Swift$ and $ROSAT$ observations spanning more than 20 years, we have obtained the first measure of the spin-period derivative of the X-ray pulsar in the \hr\   binary system.  Although the measured spin-up rate of  $2.15\times10^{-15}$ s s$^{-1}$ is rather large for an accreting WD,   the lack of a precise distance measurement does not allow us to completely rule out this possibility and safely  identify the compact object  in this system with a NS.   

The WD interpretation requires the presence of an accretion disk. This implies  that the hot subdwarf is (nearly) filling its Roche-lobe  and/or  that the distance and wind properties of \hd\ are different from those derived from the optical/UV observations. 

However, also the possibility that the compact companion of \hd\ is a NS  is not without problems. 
If confirmed, it would indicate a binary with properties very different from those of all the other known neutron star X-ray binaries. This could be related to the peculiar nature of the mass-donor star in this system, characterized by a stellar wind much weaker and with a different composition than those of ordinary OB stars.

 A precise measurement of the distance, as will be obtained with the  $GAIA$ astrometric satellite, will make it possible to assess the nature of the compact object in \hr .   For example, a WD would be strongly disfavoured    if the distance is smaller than $\sim$4 kpc. In the meantime, it is worth to  investigate better the properties of \hd\ by applying state-of-the-art atmospheric and wind models to multiwavelength spectroscopic data of this unique star.

\section*{Acknowledgments}
This work,  supported  through financial contributions from the agreement ASI/INAF I/037/12/0 and from PRIN INAF 2014, is based on data from observations with {\it XMM-Newton}, an ESA science mission with instruments and contributions directly funded by ESA Member States and the USA (NASA).

\bibliographystyle{aa}   
\bibliography{biblio_HD.bib}

\bsp

\label{lastpage}

\end{document}